# Laboratory and tentative interstellar detection of *trans*-methyl formate using the publicly available Green Bank Telescope PRIMOS survey

Justin L. Neill,[1,2] Matt T. Muckle,[1] Daniel P. Zaleski,[1] Amanda L. Steber,[1] Brooks H. Pate,[1] Valerio Lattanzi,[3] Silvia Spezzano,[3] Michael C. McCarthy,[3] and Anthony J. Remijan[4]

[1]Department of Chemistry, University of Virginia, McCormick Rd., P.O. Box 400319, Charlottesville, VA 22904; brookspate@virginia.edu

[2]Department of Astronomy, University of Michigan, 500 Church St., Ann Arbor, MI 48109; jneill@umich.edu

[3]Harvard-Smithsonian Center for Astrophysics, 60 Garden St., Cambridge, MA 02138, and School of Engineering and Applied Sciences, Harvard University, 29 Oxford St., Cambridge, MA 02138

[4]National Radio Astronomy Observatory, 520 Edgemont Rd., Charlottesville, VA 22903; aremijan@nrao.edu

Running title: Detection of trans-Methyl Formate

## Abstract

The rotational spectrum of the higher-energy *trans* conformational isomer of methyl formate has been assigned for the first time using several pulsed-jet Fourier transform microwave spectrometers in the 6-60 GHz frequency range. This species has also been sought toward the Sagittarius B2(N) molecular cloud using the publicly available PRIMOS survey from the Green Bank Telescope. We detect seven absorption features in the survey that coincide with laboratory transitions of *trans*-methyl formate, from which we derive a column density of 3.1 (+2.6, -1.2) × $10^{13}$ $cm^{-2}$ and a rotational temperature of 7.6 ± 1.5 K. This excitation temperature is significantly lower than that of the more stable *cis* conformer in the same source but is consistent with that of other complex molecular species recently detected in Sgr B2(N). The difference in the rotational temperatures of the two conformers suggests that they have different spatial distributions in this source. As the abundance of *trans*-methyl formate is far higher than would be expected if the *cis* and *trans* conformers are in thermodynamic equilibrium, processes that could preferentially form *trans*-methyl formate in this region are discussed. We also discuss measurements that could be performed to make this detection more certain. This manuscript demonstrates how publicly available broadband radio astronomical surveys of chemically rich molecular clouds can be used in conjunction with laboratory rotational spectroscopy to search for new molecules in the interstellar medium.

## 1. INTRODUCTION

Methyl formate ($HCOOCH_3$) is an abundant molecule in interstellar clouds, and has been detected, in its most stable (*cis*) isomeric form, in a number of warm star-forming regions of the interstellar medium (see *e.g.* Millar et al. 1997, Nummelin et al. 2000 and references therein). The abundance of methyl formate compared to that of its structural isomers acetic acid ($CH_3COOH$) and glycolaldehyde ($HOCH_2CHO$) is especially surprising, and has proven to be difficult to explain in models of interstellar chemistry. Of the three isomers, methyl formate is consistently found with the highest abundance across a large variety of sources, despite the fact that acetic acid is the most energetically stable of the three according to quantum chemical calculations (Dickens et al. 2001).

The chemical pathways by which methyl formate is synthesized in the interstellar medium are the subject of considerable debate. Proposed gas-phase production routes have primary involved ion-molecule reactions involving methanol ($CH_3OH$) or its protonated counterpart ($[CH_3OH_2]^+$). One reaction that has received particular interest was that of protonated methanol and formaldehyde,

$$[CH_3OH_2]^+ + H_2CO \rightarrow [HC(OH)OCH_3]^+ + H_2 \qquad (1)$$

followed by a dissociative recombination reaction,

$$[HC(OH)OCH_3]^+ + e^- \rightarrow HCOOCH_3 + H \qquad (2)$$

to yield neutral methyl formate (Blake et al. 1987). However, a computational study of this reaction (Horn et al. 2004) found that reaction (1) has an activation barrier of 128 kJ mol$^{-1}$ (15,500 K) and therefore is unfeasible at realistic molecular cloud temperatures. Horn et al. also considered a number of other gas-phase reactions that could produce methyl formate, but showed that a gas-phase-only chemical model fell far short of reproducing the abundance of methyl formate typically found in hot cores (~$10^{-8}$ – $10^{-10}$ $n(H_2)$). Additionally, the efficiency of dissociative recombination in forming complex molecules such as methyl formate from their protonated counterparts has been cast into doubt by studies using the heavy ion storage ring CRYRING (Geppert et al. 2006, Hamberg et al. 2010, Vigren et al. 2010). These studies have shown that for oxygen-containing complex molecules such as methanol, dimethyl ether, and formic acid, due to the large exothermicity of the dissociative recombination process, molecular fragmentation is a major product channel.

Motivated by the shortcomings of gas-phase processes in explaining the abundances of complex organic molecules such as methyl formate, models that include processes for complex molecule synthesis on dust grains have been formulated (Sorrell 2001, Garrod & Herbst 2006, Garrod et al. 2008). In these models, major ice components such as $H_2O$, $CH_4$, $NH_3$, $H_2CO$, and $CH_3OH$ are converted into radical species by cosmic ray-induced photodissociation, which become mobile on the grain surface as the ice warms and can engage in association reactions to form larger molecules. Eventually complex molecules are evaporated into the gas phase as the ices are warmed further or by nonthermal processes such as shocks. Models including grain chemistry have generally been more successful in explaining the high abundances of complex molecules such as methyl formate that are observed in hot cores. Laboratory simulations of interstellar ices have also shown that methyl formate can be produced under realistic irradiation conditions (Bennett & Kaiser 2007, Öberg et al. 2009, Modica & Palumbo 2010). Additionally, recent detections of methyl formate and other complex molecules in cold sources, where the likelihood of extensive gas-phase processing is low, have suggested that grain synthesis followed by non-thermal desorption is the likely production route (Requena-Torres et al. 2006, Öberg et al. 2010).

With the next generation of interferometric radio astronomical facilities, chemical spatial distribution maps are expected to emerge as a way to test possible molecular production pathways. In a recent study, spatial images of methanol, methyl formate, formic acid, and dimethyl ether in the Orion Compact Ridge were presented, showing a significant depletion in the abundance of formic acid in the region of greatest methyl formate concentration, and a high degree of similarity between the distributions of methyl formate and dimethyl ether (Neill et al. 2011). A recent modeling study simulating spatial distributions in a hot core (Aikawa et al. 2008), however, calculated that formic acid and methyl formate, if predominantly formed within ices, would have similar spatial distributions, peaking in the warmest regions. Therefore, two additional gas-phase ion-molecule reactions that were not previously included in gas-grain chemical kinetics models were proposed that could explain the relative spatial distributions of formic acid, methyl formate, and dimethyl ether in the Orion Compact Ridge (Neill et al. 2011):

$$CH_3OH + [HC(OH)_2]^+ \rightarrow [HC(OH)OCH_3]^+ + H_2O \quad (3)$$

$$[CH_3OH_2]^+ + HCOOH \rightarrow [HC(OH)OCH_3]^+ + H_2O \quad (4)$$

For both reactions there are two transition state geometries, which yield different conformations in the protonated methyl formate product (*cis* or *trans* in the C-O-C-O ester dihedral angle). For reaction (3), the Fischer esterification mechanism, both pathways have an activation barrier according to *ab initio* calculations (17 kJ $mol^{-1}$ to form *cis*, 21 kJ $mol^{-1}$ to form *trans*). While these barriers are far lower than that of reaction (1), they are still high enough that its feasibility as an interstellar reaction mechanism at typical hot core temperatures (~100 K) is doubtful. By reaction (4), the methyl cation transfer reaction, which was previously proposed by Ehrenfreund and Charnley (2000), the *trans* conformer can form without an activation barrier, while the formation of the *cis* product has a barrier of 10 kJ $mol^{-1}$. Therefore, the primary product of this reaction is expected to be *trans*-protonated methyl formate, which could lead to an observable abundance of neutral *trans*-methyl formate after dissociative recombination. An analogous methyl cation transfer reaction between methanol and protonated methanol can also produce protonated dimethyl ether in the gas phase without a barrier (Bouchoux & Choret 1997); this reaction has been proposed to be an important interstellar synthesis pathway for dimethyl ether (Peeters et al. 2006).

Motivated by the prediction that *trans*-methyl formate could be produced at an observable abundance by this gas-phase process, we present the laboratory assignment of the rotational spectrum of *trans*-methyl formate. Previous spectroscopic characterizations of this species have been performed using infrared spectroscopy of low-temperature argon matrices with nonthermal population distributions of the two conformers. (Blom & Günthard 1981, Müller et al. 1983). We also report a tentative detection of *trans*-methyl formate in the Sagittarius B2(N) region using publicly available survey data from the NRAO Green Bank Telescope. The *trans* conformer is less stable than *cis* by 25 kJ $mol^{-1}$ (3000 K), so at typical temperatures of interstellar clouds, the population ratio is very large at thermal equilibrium (*e.g.*, $\sim 10^{13}$:1 *cis*:*trans* ratio at 100 K). However, a large energy barrier must be crossed to interconvert between conformers (Senent et al. 2005), as shown in Figure 1, so it is unlikely that equilibrium will be established between the two conformers in molecular clouds, and instead the relative abundances of the two species should reflect conformer-specific chemical processes. The *trans* isomer has a far lower rotational temperature than *cis* in this source, suggesting that the two conformers may have different spatial distributions and production mechanisms.

This manuscript also demonstrates a method by which new molecules predicted to be present in the interstellar medium on the basis of chemical models can be sought by the combination of new laboratory measurements and publicly available survey data. The newest generation of radio astronomical facilities is beginning to produce a massive database of broad-bandwidth surveys of chemically rich sources from the microwave to the millimeter-wave/THz spectral regions. With this extensive data archive available to the public, we anticipate that it will become routine that publications describing laboratory rotational spectroscopic studies of candidate interstellar molecules will include searches for these molecules in relevant astronomical sources. The research model we present here can improve the efficiency with which the enormous amount of chemical information present in these surveys can be extracted.

## 2. EXPERIMENT

The laboratory microwave spectrum of *trans*-methyl formate was measured on several pulsed-jet Fourier transform microwave (FTMW) spectrometers: chirped-pulse Fourier transform microwave (CP-FTMW) spectrometers operating from 7-18.5 GHz (Brown et al. 2008) and 25-40 GHz (Zaleski et al., submitted), and two cavity FTMW spectrometers based on the design of Balle and Flygare (1981), a miniature FTMW spectrometer based on a NIST design (Suenram et al. 1999) and a cryogenically cooled FTMW spectrometer (Grabow et al. 2005). A sample of 0.2% methyl formate (99%, purchased from Aldrich and used without further purification) in an 80:20 neon:helium ("first run neon") expansion was used. Additionally, due to the low population of *trans*-methyl formate at room temperature equilibrium, a pulsed discharge nozzle of the type previously described by McCarthy et al. (2000), operated with a voltage of approximately 800 V, was used to increase the population of *trans*-methyl formate in the supersonic jet by a factor of approximately 30. Laboratory spectra using CP-FTMW spectrometers are shown in Figure 2.

Due to the strong internal rotor effects present in the spectrum of *trans*-methyl formate, FTMW-MW double resonance measurements played a crucial role in confirming the spectral assignments reported here and in locating new transitions. The use of an FTMW cavity spectrometer for double resonance measurements has been previously described (Nakajima et al. 2002; Suma et al. 2004; Douglass et al. 2006). In this experiment, the FTMW cavity is tuned to a rotational transition. Immediately after the FTMW polarizing pulse, a second microwave pulse, typically with duration 3 μs, is applied. If the double resonance microwave pulse is resonant with a transition that shares a quantum state with the FTMW-monitored transition, the prepared coherence is destroyed, reducing the intensity of the measured signal typically by more than 80%. Double resonance sources with at least 100 mW of power from 6-60 GHz were employed. The double resonance frequency source could either be programmed to test for connectivity against a list of observed transitions, or to scan through a frequency range.

**3. OBSERVATIONS**

Observations of the *cis* and *trans* conformers of methyl formate were conducted as part of the NRAO 100 m Robert C. Byrd Green Bank Telescope (GBT) PRebiotic Interstellar MOlecule Survey (PRIMOS) Legacy Project. This NRAO key project started in 2008 January and concluded in 2011 July. The PRIMOS project recorded a nearly frequency continuous astronomical spectrum from the Sgr B2(N) molecular cloud (J2000 pointing position of $\alpha = 17^h47^m19^s.8$, $\delta = -28°22'17''$) between 1 and 50 GHz. The source was observed above 10° elevation from source rise to source set, when possible. An LSR source velocity of +64 km $s^{-1}$ was assumed. The GBT spectrometer was configured in its eight intermediate-frequency (IF), 200 MHz three-level mode, which enabled the observation of four 200 MHz frequency bands at a time in two polarizations through the use of offset oscillators in the IF. Antenna temperatures were recorded on the $T_a^*$ scale (Ulich & Haas 1976) with estimated 20% uncertainties. Data were taken in the OFF-ON position-switching mode, with the OFF position 60' east in azimuth with respect to the ON-source position. A single scan consisted of 2 minutes in the OFF-source position followed by 2 minutes in the ON-source position. Automatically updated dynamic pointing and focusing corrections were employed based on real-time temperature measurements of the structure input to a thermal model of the GBT; zero points were typically adjusted every 2 hours or less using the quasar 1733-130 for calibration.

The processes and procedures for the data reduction we followed to reduce and analyze the data using the GBTIDL package can be found online.[1] Before any profile fitting was performed, the continuum level was subtracted from each spectrum, using up to a third-order polynomial fit to the baseline. This removed any instrumental slopes in the bandpass so that baselines would be flat enough for a profile analysis, but the shape of the original line would be preserved. The two polarization outputs from the spectrometer were averaged in the final data reduction process to improve the signal-to-noise ratio, except for in the $K_a$ band (26-40 GHz), where the receiver was limited to one polarization for these observations. Access to the entire PRIMOS dataset and specifics on the observing strategy, including the overall frequency coverage, are available at http://www.cv.nrao.edu/~aremijan/PRIMOS/.

[1] http://gbtidl.nrao.edu; http://www.gb.nrao.edu/GBT/DA/gbtidl/users_guide.2pt8.pdf; http://www.gb.nrao.edu/GBT/DA/gbtidl/gbtidl_calibration.pdf

## 4. LABORATORY SPECTRUM OF *TRANS*-METHYL FORMATE

Like that of *cis*-methyl formate, the rotational spectrum of *trans*-methyl formate is complicated by internal methyl rotation. A total of 33 transitions of the dominant ($H^{12}C^{16}O^{16}O^{12}CH_3$) isotopologue of *trans*-methyl formate were detected in the laboratory in this study, 24 of the *A* symmetry (nondegenerate) species and 9 of the *E* symmetry (degenerate) species, with *a*- and *b*-type selection rules. In addition, transitions of each of the two singly-substituted $^{13}C$ isotopologues were observed in natural abundance. All transitions up to 40 GHz with appreciable intensity at pulsed-jet temperatures ($T_{rot}$ ~ 1-2 K) have been measured. The standard asymmetric top quantum numbers $J_{KaKc}$ are used to designate energy levels for the *A* symmetry species, but due to strong coupling between the internal rotation of the methyl group and the overall rotation of the molecule, the symmetric rotor quantum numbers $J_K$, with a signed value of *K*, are used for the *E* symmetry species. The frequencies of all measured transitions can be found in Table 1.

The literature devoted to the analysis of rotational spectra of asymmetric top molecules with low-barrier methyl internal rotors is extensive (Kleiner 2010). Several free programs are available to fit these spectra;[2] for this study we used the rho-axis method (RAM) as implemented in the BELGI program (Hougen et al. 1994). As the supersonic expansion cools out all torsionally excited states in these spectrometers, there is some indeterminacy in the Hamlitonian parameters presented here. In particular, when fitting only the ground torsional state, there is a strong correlation between the internal rotation constant *F* and the methyl rotation barrier height $V_3$; therefore, we fixed *F* to its *ab initio* value. The derived rotation and internal-rotation parameters that result from this fit are presented in Table 2, compared to values calculated by electronic structure theory using the Gaussian 09 software package (Frisch et al. 2009).

[2]Kisiel, Z. Programs for ROtational SPEctroscopy, available at http://www.ifpan.edu.pl/~kisiel/prospe.htm.

From the derived rotational constants of the three isotopologues, substitution coordinates for the two carbon atoms in *trans*-methyl formate can be derived. The RAM inverse inertial tensor was diagonalized to derive rotational constants in the principal axis system, from which the equations of Kraitchman (1953) were applied to derive the positions of the carbon atoms in the principal axis system (Gordy & Cook 1984). In Table 3, these substitution coordinates are compared to the *ab initio* calculated positions of the two carbon atoms. The excellent agreement serves as further confirmation of the molecular identity of the assigned spectrum.

From the relative intensities of *cis*- and *trans*-methyl formate transitions in the CP-FTMW spectrometer without the pulsed discharge nozzle applied, the energy difference between the two conformers can be estimated if it is assumed that no conformational cooling occurs in the supersonic expansion. As Ruoff et al. (1990) showed, conformational relaxation typically only occurs in supersonic jets when the barrier to relaxation is ~5 kJ mol$^{-1}$ or less, so it is likely to be negligible for methyl formate where the barrier for *trans* to isomerize to *cis* is calculated to be 35 kJ mol$^{-1}$. We assume quadratic scaling of transition intensity with dipole moment, and use values of $\mu_a$ = 1.6 D for the *cis* conformer (Curl 1959) and $\mu_a$ = 4.2 D for the *trans* conformer (calculated at an MP2/6-311++G(d,p) level of theory). By comparison of the $1_{01}$-$0_{00}$ *A* species transitions of the two conformers, we estimate a population ratio of 22,000:1 (*cis*:*trans*) at room temperature, which corresponds to an energy difference of 24.6 kJ mol$^{-1}$, or 2980 K. This is in excellent agreement with electronic structure calculations, both by Senent et al. (2005) and in this work.

The threefold methyl potential barrier $V_3$ of *trans*-methyl formate determined in this study (12.955(17) $cm^{-1}$) is considerably lower than that of *cis*-methyl formate (379.4 $cm^{-1}$) (Karakawa et al. 2001). While this report presents the first experimental determination of this barrier, it has been calculated previously using electronic structure theory. Wiberg et al. (2005) compared the methyl rotation barriers of *cis*- and *trans*-methyl formate to those of structurally related molecules, and concluded that the methyl $V_3$ barrier observed for *cis*-methyl formate is due to repulsion between the carbonyl oxygen atom and the methyl proton when the two are eclipsed, leading to a strong preference for a staggered geometry (see Figure 1). In the *trans* form, however, the carbonyl oxygen is not in close proximity to the methyl protons, and this effect vanishes, resulting in a very small barrier to methyl internal rotation.

## 5. METHYL FORMATE OBSERVATIONS TOWARDS SGR B2(N)

### *5.1. Tentative Detection of* trans-*Methyl Formate*

Using the laboratory transition frequencies reported above, *trans*-methyl formate was sought in Sgr B2(N). Because of the dipole moment components of this species (calculated $\mu_a$ = 4.2 D, $\mu_b$ = 2.5 D at a MP2/6-311++G(d,p) level of theory), the strongest transitions in the frequency range of the GBT are expected to be *a*-type transitions with $J'$ = 1-5, $K_a$ = 0, 1, as these transitions have the highest line strengths and lowest energies. Table 4 presents relevant parameters for the transitions that were sought in this data set. In Figure 3 we show an overlay of the laboratory spectrum from Figure 2 with a portion of the Sgr B2(N) PRIMOS spectrum in the region of the $J$ = 1-0 transitions of the *A* and *E* species. Both of the observed transitions in the laboratory correspond closely to absorption features of equal intensity (-10 mK). These two transitions, as can be seen from Figures 2 and 3, do not have equal intensities in the laboratory spectrum, with the $A$ species transition being more intense; we attribute this to the $E$ species population cooling into the $K = 1$ ladder (which is lower in energy than the $K = 0$ ladder due to internal rotation interactions) in the cold ($T_{rot} \sim 2$ K) supersonic jet. However, at higher temperatures than this, we expect the $A$ and $E$ transitions to have equal intensity, as is observed for the two candidate features in the Sgr B2(N) survey.

Absorption features are also observed at the frequencies of the $J$ = 2-1 and 3-2, $K_a = 0$ transitions, for both $A$ and $E$ symmetry species. In Figure 4 we show the $J$ = 1-0, 2-1, and 3-2 transition pairs. None of these transitions are attributable to other known interstellar molecules. The $A$-$E$ splittings vary significantly between these three pairs of transitions (83.21 MHz, 120.81 MHz, and 73.94 MHz for $J$ = 1-0, 2-1, and 3-2, respectively), so the identification of each of these transition pairs with equal intensities and LSR velocities is suggestive that these features are due to *trans*-methyl formate. As many of the lines in the PRIMOS survey are unassigned, a full spectral line catalog of the PRIMOS survey between 18-26 GHz has been made to estimate statistically the likelihood of coincidental assignments. In this 8 GHz range, a total of 330 transitions are observed (both absorption and emission, including assigned and unassigned features) with an intensity of 10 mK or greater; the typical noise level is ~5-10 mK in this frequency range. Recombination lines are excluded from this count, as they are readily distinguished by their spectral line shape. Of these transitions, a total of 191 of these transitions are absorption features, of which 132 have intensities between -10 and -30 mK (as the six transitions in Figure 4). We use this spectral region as a proxy for the spectral line density in the PRIMOS survey as a whole, and note that there is not a significant increase in spectral line density in the other frequency ranges (9 and 27 GHz) where we report *trans*-methyl formate features.

Most molecular transitions in the PRIMOS survey have full-width at half-maximum (FWHM) linewidths of at most 25 km $s^{-1}$, which corresponds to 1.5 MHz at 18 GHz and 2.1 MHz at 26 GHz. We therefore assess the probability of a transition lying within one FWHM of a given frequency, which we approximate as a 4 MHz window (± 2 MHz). (Note that all of the detected lines presented in Table 4 are significantly closer than this condition, within 7 km $s^{-1}$ of the primary velocity of the Sgr B2(N) source, 64 km $s^{-1}$.) Using the line count above, there is a probability of 0.152 that at least one transition of any intensity, absorption or emission, will fall within a given 4 MHz window, a probability of 0.092 for an absorption feature, and a probability of 0.064 for an absorption feature with an intensity of between -10 and -30 mK. We present six detected features in Figure 4; the odds of features being found within 2 MHz of each of these frequencies are 80,000:1 (for any transition), 1,700,000:1 (for an absorption transition of any intensity), and 14,700,000:1 (for absorption features between -10 and -30 mK). These probabilities of course are strongly dependent on the spectral line density of the survey. For comparison, the 80-116 GHz survey of Sgr B2(N) by Belloche et al. (2008, 2009) using the IRAM 30 m telescope has a quoted line density of about 100 features per GHz. As the PRIMOS survey between 18-26 GHz contains about 40 features per GHz, and transitions are ~4 times narrower in frequency (assuming the same width in velocity units), the total spectral line density, in terms of the fraction of spectral channels occupied by a line, is approximately an order of magnitude lower in the PRIMOS survey. That is, while there is a 0.152 probability of a line being found within a two-FWHM window in the PRIMOS survey, the probability of finding at least one line in a similar window in the IRAM 30 m survey (assuming again a 25 km $s^{-1}$ feature width, or ~8 MHz at 100 GHz) in the 3 mm spectral window is ~0.8.

Using this analysis as grounds to suggest that the observed transitions are likely due to *trans*-methyl formate, we derive an excitation temperature and column density for this species. An absorption feature has also been detected at the frequency of the $J_{KaKc}$ = $2_{11}$-$1_{10}$ transition of the *A* symmetry species. Because of the large energy level shifts in the *E* species due to low-barrier methyl internal rotation, the corresponding transition ($J_K$ = $2_{-1}$-$1_{-1}$) has a higher predicted lower-state energy ($E_l$ = 10 K) than can be populated at supersonic expansion temperatures, so this transition was not measured in the laboratory. Because the $2_{11}$-$1_{10}$ (*A*) transition is close in frequency to the $2_{02}$-$1_{01}$ (*A*) transition, the GBT beamwidth is nearly the same for the two transitions and so samples the same spatial region. However, since the difference in the lower-state energies of the two transitions is reasonably large, the intensity ratio between them depends strongly on excitation temperature. The GBT observations of these two transitions are shown in panel (A) of Figure 5, while the calculated intensity ratio between the two transitions as a function of excitation temperature (assuming local thermodynamic equilibrium (LTE) level populations) is shown in panel (B). Based on the observed relative intensity ratio of 0.56 ± 0.05, we derive an excitation temperature of 7.6 ± 1.5 K.

Using this temperature, a beam-averaged column density for *trans*-methyl formate was derived, using the equation for absorption features from Remijan et al. (2005):

$$N_T = 8.5\times10^9 \frac{Q_r(\Delta T_a^* \Delta V / \eta_B)}{(T_{ex} - T_c/\eta_B) S\mu^2 (e^{-E_l/T_{ex}} - e^{-E_u/T_{ex}})} \mathrm{cm}^{-2} \quad (5)$$

In this equation, $Q_r$ is the rotational partition function; $\Delta T_a^* \Delta V$ is the product of the antenna temperature of the transition (in mK) and the full-width at half-maximum width (in km s$^{-1}$); $\eta_B$ is the telescope beam efficiency; $T_{ex}$ is the excitation temperature; $T_c$ is the source continuum temperature; $S\mu^2$ is the product of the intrinsic transition line strength and the square of the *a*-type dipole moment component in D$^2$; and $E_l$ and $E_u$ are the energies of the lower and upper rotational levels of the transition. Due to the different energy level structures of the *A* and *E* species, column densities were determined separately for the two torsional subspecies. For the *A* species, a value of $Q_r$ = 4.67 $T_{ex}^{1.5}$ was used, derived from an effective rigid-rotor fit to this species while for the *E* species a direct state count was performed using predicted energy levels, yielding a value of $Q_r$ = 108.3 at 7.6 K. A value of $N_T$ was then derived for each detected transition, and the values averaged to yield a final estimate. From this we derive an average column density of 1.2 (+1.2, -0.6) × $10^{13}$ $cm^{-2}$ for the *A* species, and 1.9 (+2.3, -1.0) × $10^{13}$ $cm^{-2}$ for the *E* species, for a total column density of 3.1 (+2.6, -1.2) × $10^{13}$ $cm^{-2}$. The column densities are the same for the *A* and *E* species within the measurement error.

In addition to the primary LSR velocity of +64 km $s^{-1}$, a number of molecules in Sgr B2(N) have velocity components near +73 and/or +82 km $s^{-1}$, including glycolaldehyde (Hollis et al. 2004a) and several cyanides and isocyanides (Remijan et al. 2005). While the signal-to-noise ratio on individual transitions of *trans*-methyl formate is low, several transitions show an asymmetry in the lineshape that suggests an absorption component near these velocities. In Figure 6, the average of six transitions (all confidently detected transitions except the $3_{03}$-$2_{02}$ transition of the *A* species, which is omitted due to a strong nearby line) is presented, showing clearer evidence for a higher-velocity component.

A number of other transitions of *trans*-methyl formate have been sought toward Sgr B2(N) beyond those discussed thus far. Four transitions ($2_{12}$-$1_{11}$ (*A*), $3_{13}$-$2_{12}$ (*A*), $2_1$-$1_1$ (*E*), and $3_1$-$2_1$ (*E*)) do not have clearly detected features, but the noise levels in the observations at these frequencies are comparable to or higher than the predicted intensities expected based on the temperature and column density parameters derived above, and tentative evidence for absorption features can be seen. GBT observations of these four transitions are shown in Figure 7. For the $J$ = 4-3 transitions near 36 GHz (see Table 1), assuming $T_{ex}$ = 7.6 K, little flux is expected because $T_{ex}$ is nearly equal to the observed continuum temperature corrected for beam efficiency ($T_c/\eta_B$); however, the intensity of this line, and whether it is expected to be in emission or absorption, is highly dependent on $T_{ex}$. The $4_{04}$-$3_{03}$ (A) transition is obscured by a strong (transition of $HC_3N$, while the $4_0$-$3_0$ ($E$) transition is not detected, with an rms noise level of ~15 mK. The $5_{05}$-$4_{04}$ ($A$) transition, which was not measured in the laboratory but is calculated to have a frequency of 45593.037(3) MHz, was searched for, with no detection at an 8 mK rms noise level. Because of the lower source continuum temperature ($T_c$ = 2.66 K) at this frequency, this line would be expected to be in emission, with an antenna temperature of 35 mK, based on the column density and excitation temperature derived above; however, this line intensity is also highly sensitive to the parameters of our simple excitation model. In addition, for these higher-frequency lines, the half-power beamwidth of the GBT is smaller (21″ at 36 GHz, 16″ at 45 GHz) than in the lower-frequency lines, and so the telescope beam is probing a different spatial region within Sgr B2(N).

In summary, we present seven transitions of *trans*-methyl formate where clear absorption features are found in this survey, as well as four other tentative features, and show that the likelihood of this being coincidental is small. However, as there is significant uncertainty in the excitation model we present, further measurements could provide additional credence to this detection. In addition to the four marginally detected lines presented in Figure 7, which could be the target of deeper integrations with the GBT, there are a number of other transitions in Table 1 that are expected to be detectable in Sgr B2(N) given sufficient integration time (well beyond that of the PRIMOS survey); these transitions include higher-$K$ $a$-type transitions, as well as $b$-type transitions, that have lower line strengths than the $K$ = 0 transitions we present here.

Additionally, observations of higher-frequency transitions of *trans*-methyl formate could be used to probe any warmer components that might exist. Millimeter observations, however, will first require laboratory millimeter- and submillimeter-wave spectroscopic measurements. These experiments will be difficult on a room-temperature sample, as is used for many spectrometers in this frequency range, because the equilibrium population of *trans*-methyl formate at room temperature is lower than that of several isotopomers ($^{13}C$, $^{18}O$) as well as an estimated 50 vibrationally excited species of *cis*-methyl formate, which could lead to confusion limit issues. Additionally, particularly for the *E* symmetry species for which the energy level structure is highly perturbed, extrapolation of the measured transitions to millimeter-wave frequencies is likely to result in significant error. An efficient discharge method for producing a large abundance of *trans*-methyl formate for millimeter-wave study, similar to that employed in this study at microwave frequencies, will likely be needed.

### *5.2. Comparison to* cis-*Methyl Formate*

A number of determinations of the rotational temperature and column density of *cis*-methyl formate in Sgr B2(N) have been made through analyses of its millimeter-wave spectrum (Cummins et al. 1986, Turner 1991, Nummelin et al. 2000, Belloche et al. 2009). Interferometric studies have revealed that this molecule is predominantly found in the compact (≤5″) Large Molecule Heimat (LMH) hot core (Mehringer et al. 1997, Hollis et al. 2001, Belloche et al. 2008). In a recent study using the IRAM 30-m millimeter telescope (Belloche et al. 2009), assuming a source size of 4″, a rotational temperature of 80 K and a column density of $4.5 \times 10^{17}$ cm$^{-2}$ was derived. Over 160 transitions of *cis*-methyl formate were searched for in the PRIMOS survey between 1.5 and 48 GHz, of which 66 spectral features were detected. Nearly all transitions of *cis*-methyl formate detected in the PRIMOS survey are found in emission, while all observed features of *trans* are in absorption, which suggests that *cis*-methyl formate is warmer than *trans* in this source. However, due to opacity effects and the ambiguity of the overall distribution of the emission, a temperature and column density estimate for *cis*-methyl formate has not yet been determined from these data, but the analysis is ongoing. An estimate of the *cis*:*trans* abundance ratio in the GBT observations therefore cannot be provided here, and is not meaningful if the two conformers are located in different spatial regions. However, because of the large energy difference between these two conformers, it is certain that based on the column density we have tentatively derived, *trans*-methyl formate is found at a far higher relative abundance than would be present if the two conformers were in thermal equilibrium at the kinetic temperature of the cloud.

## 6. DISCUSSION

The tentative detection of *trans*-methyl formate is one of the first cases where more than one conformational isomer of a molecule has been detected in the interstellar medium; this is partially due to the fact that larger molecules, which are more likely to have multiple conformers, are more difficult to observe because these molecules have larger rotational partition functions and typically have lower abundances. Nevertheless, there are two molecules, ethanol and vinyl alcohol, for which detections of multiple conformers have been reported. Both conformations (*trans* and *gauche*) of ethanol have been detected in the Orion KL region. (Pearson et al. 1997) Because the energy difference between these two conformers (57 K) and the barrier to interconversion (560 K) are small (Pearson et al. 1997), the higher-energy *gauche* conformer is likely to be thermally excited at hot core temperatures. For vinyl alcohol ($CH_2CHOH$), the *syn* and *anti* conformers have both been detected in Sgr B2(N) (Turner & Apponi 2001). Due to the small number of observed lines in this study (five for the *anti* conformer, two for *syn*), the temperatures and column densities of the two conformers could not be independently determined. However, the large energy difference between these conformers (1100 K) suggests that an observable abundance of the higher-energy *anti* conformer could be due to kinetic effects.

The different physical characteristics of *cis* and *trans*-methyl formate suggest that the two conformers are not cospatially located, and could be produced by different mechanisms. Based on its rotational temperature, *trans*-methyl formate is likely to be predominantly in the colder region surrounding the LMH hot core. A number of other complex organic molecules have been found to contain components with low rotational temperatures, including glycolaldehyde ($CH_2OHCHO$) (Hollis et al. 2004a), formamide ($HCONH_2$) (Halfen et al. 2011), acetamide ($CH_3CONH_2$) (Hollis et al. 2006, Halfen et al. 2011), propenal ($CH_2CHCHO$) (Hollis et al. 2004b), and propanal ($CH_3CH_2CHO$) (Hollis et al. 2004b). Interferometric observations of glycolaldehyde (Hollis et al. 2001) and acetaldehyde ($CH_3CHO$) (Chengalur & Kanekar 2003) have revealed that these two molecules, in contrast to *cis*-methyl formate and some other complex molecules, are spread over a region that is ≥60″ in extent. The presence of complex gas-phase molecules in this region has been attributed to shocks releasing material from grain mantles, as inferred from the presence of a large number of masers (Chengalur & Kanekar 2003, Sato et al. 2000, Mehringer & Menten 1997). Therefore, we consider production mechanisms by which *trans*-methyl formate could be produced at a non-thermal abundance in this region.

First we consider the possibility that grain chemistry could yield this result. The conformer ratio produced by grain radical-radical combination reactions as in the model of Garrod et al. (2008) is not known experimentally, but the exothermicity of this reaction is far greater than the barrier to isomerization between conformers, which could allow for the production of significant quantities of *trans*-methyl formate. Once *trans*-methyl formate is synthesized, its re-conversion to the more stable *cis* isomer could be slow within the grain, particularly if it is liberated into the gas phase at low temperatures through shocks. Further experiments are required to determine whether grain synthesis could explain the observed gas-phase abundance of *trans*-methyl formate.

Alternatively, the observed *trans*-methyl formate could be the result of gas-phase processing. As discussed by Neill et al. (2011), the gas-phase methyl cation transfer reaction (4) is expected, based on electronic structure calculations, to lead to the production of a significant abundance of *trans*-methyl formate. Because the reaction to form the *trans* isomer of protonated methyl formate does not have an activation barrier, it could have a significant rate coefficient at low temperatures if the reactants, formic acid and protonated methanol, are present in the gas phase. In a chemical modeling study by Laas et al. (2011), formic acid and methanol are formed on grains at low temperatures, then thermally desorbed into the gas phase at ~100 K. Reasonable abundances of protonated methanol can be produced by the reaction of methanol with $H_3^+$ (Huntress & Bowers 1973, Fiaux et al. 1976), so *trans*-methyl formate is produced in the model of Laas et al. through reaction (4). If non-thermal desorption were instead to release methanol and formic acid into the gas phase at lower temperatures, then methanol protonation, followed by the synthesis of *trans*-methyl formate, might occur efficiently.

Another gas-phase process to produce methyl formate that could proceed at low temperatures is a neutral-neutral radiative association reaction (Barker 1992), *e.g.*

$$HCO + OCH_3 \rightleftarrows (HCOOCH_3)^* \rightarrow HCOOCH_3 + h\nu \qquad (6)$$

where two radicals combine to form vibrationally excited methyl formate; at the densities typical for interstellar clouds, the dominant relaxation mechanism for association reactions is through emitting infrared radiation (Herbst & Klemperer 1973). This excited species, in addition to dissociating back into the reactants, could also undergo a unimolecular dissociation process, leading to $CH_3OH$ + CO or $H_2CO$ + $H_2CO$; these processes have activation barriers of about 300 kJ $mol^{-1}$ (Francisco 2003, Metcalfe et al. 2010), considerably less than the 450 kJ $mol^{-1}$ released in the association reaction (Glockler 1958). Therefore, the excited methyl formate molecule would have to emit multiple infrared photons before the product molecule is stable against dissociation, which could make the product yield of reaction (6) low. Additionally, because the reactants in equation (6) are both neutral, the collision cross sections would be lower than for an ion-neutral reaction, and so the overall reaction rate would be lower given equal densities. The conformational distribution that would result from this reaction is not known, but since the excited intermediate contains far more internal energy than is required for conformational isomerization, it would be likely to form a non-equilibrium mix of *cis* and *trans*. Experimental and/or theoretical work will be needed to fully assess the viability of this reaction.

Finally, we consider gas-phase processes that can convert *cis*-methyl formate to *trans* or vice versa. Because of the large energy barrier of a direct isomerization process (as seen in Figure 1), unimolecular isomerization is likely extremely slow. However, an indirect isomerization process could occur if methyl formate were protonated by an abundant molecular ion such as $H_3^+$, $HCO^+$, or $H_3O^+$, followed by dissociative recombination (reaction (2)). Both reactions have exothermicities, based on gas-phase proton affinities (Hunter & Lias 1998), that exceed the barrier to isomerization between the two conformers of neutral methyl formate; the corresponding barriers in protonated methyl formate are calculated to be similar. Therefore, isomerization could occur as part of either step as the highly excited product molecule relaxes. These two reactions could form a cycle that drives the conformer abundance ratio toward a value that is likely to differ considerably from the thermal equilibrium value.

It should also be noted that formic acid has a similar conformational potential energy surface to that of methyl formate, with two analogous conformational isomers. The energy difference between the two conformers of formic acid is 16 kJ $mol^{-1}$ (with the more stable conformer referred to in the literature as *trans*[3]), while unimolecular isomerization from *cis* to *trans* is calculated to have a barrier of ~35 kJ $mol^{-1}$ (Goddard et al. 1992). In considering processes that could lead to the observed abundance of *trans*-methyl formate, therefore, the relative abundances of *cis* and *trans*-formic acid produced by similar chemistry should also be considered. To date, all interstellar detections of formic acid have been of the more energetically stable conformer, despite the fact that the rotational spectrum of the less stable *cis* form is well known up to THz frequencies (Hocking 1976, Winnewisser et al. 2002). This nondetection should therefore be considered as a constraint in production mechanisms. If *trans*-methyl formate is produced through isomerization during the protonation/dissociative recombination reaction sequence as described above, formic acid is likely to undergo this chemistry as well, and a significant abundance of *cis*-formic acid would therefore be expected in the same region. There is, however, one noteworthy difference: because isomerization in formic acid involves the motion of a single H atom rather than of a methyl group, tunneling could cause the unimolecular interconversion between neutral formic acid conformers to be faster than for methyl formate,

[3] The conformers of formic acid are typically described by the stereochemical relationship of the two H atoms across the C-O bond, whereas conformers of methyl formate are described by the relationship of –O and $–CH_3$. Therefore the nomenclature is reversed between the two molecules; *cis*-methyl formate and *trans*-formic acid are structurally analogous, and vice versa. Here we maintain the literature nomenclature for both molecules, to minimize confusion.

which would drive the conformer abundances of formic acid back towards thermodynamic equilibrium.

## 7. CONCLUSION

We have presented the first laboratory rotational spectroscopic characterization of the higher-energy *trans* conformational isomer of methyl formate, along with a tentative detection in the Sagittarius B2(N) molecular cloud, using publicly available survey data. Because of the large energy difference between the two conformers, any detection of *trans*-methyl formate in an interstellar cloud is an indicator of non-equilibrium chemistry between the two conformers, and we have discussed processes that could be responsible for this. The lower rotational temperature of *trans* compared to that of *cis* in the same source indicates that the two conformers have different spatial distributions and may be formed by different processes. High-resolution interferometric observations of both the *cis* and *trans* methyl formate isomers will be essential in fully characterizing the degree to which their spatial distributions are correlated, as well as to understand the physical conditions in the regions where *trans*-methyl formate could be produced.

The authors acknowledge support from the NSF Centers for Chemical Innovation (CHE-0847919). We also thank Eric Herbst and Susanna Widicus Weaver for fruitful discussions, and the anonymous referee for helpful comments to improve the manuscript. The National Radio Astronomy Observatory is a facility of the National Science Foundation operated under cooperative agreement by Associated Universities, Inc.

**Table 1**. Rotational transitions of *trans*-methyl formate and its two singly-substituted $^{13}C$ isotopologues observed in the laboratory.

*A* symmetry species

| $J'$ | $K_a'$ | $K_c'$ | $J''$ | $K_a''$ | $K_c''$ | $H^{12}COO^{12}CH_3$ Frequency (MHz) | $H^{13}COO^{12}CH_3$ Frequency (MHz) | $H^{12}COO^{13}CH_3$ Frequency (MHz) |
|---|---|---|---|---|---|---|---|---|
| 7 | 0 | 7 | 6 | 1 | 6 | 8812.533(3) | -- | -- |
| 1 | 0 | 1 | 0 | 0 | 0 | 9124.221(3) | 9091.497(5) | 8878.750(5) |
| 7 | 1 | 6 | 7 | 1 | 7 | 9160.378(3) | -- | -- |
| 4 | 1 | 4 | 5 | 0 | 5 | 11124.535(3) | -- | -- |
| 8 | 1 | 7 | 8 | 1 | 8 | 11776.682(3) | -- | -- |
| 9 | 1 | 8 | 9 | 1 | 9 | 14719.245(3) | -- | -- |
| 2 | 1 | 2 | 1 | 1 | 1 | 17921.508(3) | -- | -- |
| 10 | 1 | 9 | 10 | 1 | 10 | 17987.583(3) | -- | -- |
| 2 | 0 | 2 | 1 | 0 | 1 | 18247.038(3) | 18181.567(5) | 17756.242(5) |
| 2 | 1 | 1 | 1 | 1 | 0 | 18575.916(3) | -- | 18068.413(5) |
| 3 | 1 | 3 | 4 | 0 | 4 | 20877.496(3) | -- | -- |
| 3 | 1 | 3 | 2 | 1 | 2 | 26881.342(3) | 26779.167(10) | 26169.865(10) |
| 3 | 0 | 3 | 2 | 0 | 2 | 27367.048(3) | 27268.729(10) | 26631.176(10) |
| 3 | 1 | 2 | 2 | 1 | 1 | 27862.941(3) | 27768.743(10) | 27101.760(10) |
| 2 | 1 | 2 | 3 | 0 | 3 | 30479.000(3) | -- | -- |
| 4 | 1 | 4 | 3 | 1 | 3 | 35840.078(3) | 35703.788(10) | 34891.610(10) |
| 4 | 0 | 4 | 3 | 0 | 3 | 36482.850(3) | 36351.539(10) | 35502.325(10) |
| 4 | 1 | 3 | 3 | 1 | 2 | 37148.834(3) | 37023.179(10) | 36134.080(10) |
| 1 | 1 | 1 | 2 | 0 | 2 | 39924.543(3) | -- | -- |
| 1 | 1 | 0 | 1 | 0 | 1 | 58498.725(50) | -- | -- |
| 2 | 1 | 1 | 2 | 0 | 2 | 58827.710(50) | -- | -- |
| 3 | 1 | 2 | 3 | 0 | 3 | 59323.533(50) | -- | -- |
| 4 | 1 | 3 | 4 | 0 | 4 | 59989.491(50) | -- | -- |
| 5 | 1 | 4 | 5 | 0 | 5 | 60829.684(50) | -- | -- |

*E* symmetry species

| $J'$ | $K'$ | $J''$ | $K''$ | $H^{12}COO^{12}CH_3$ Frequency (MHz) | $H^{13}COO^{12}CH_3$ Frequency (MHz) | $H^{12}COO^{13}CH_3$ Frequency (MHz) |
|---|---|---|---|---|---|---|
| 1 | 0 | 0 | 0 | 9207.427(3) | -- | -- |
| 2 | 1 | 1 | 1 | 17820.962(3) | -- | -- |
| 2 | 0 | 1 | 0 | 18367.848(3) | 18303.082(5) | 17872.392(5) |
| 2 | 0 | 3 | 1 | 26053.738(3) | -- | -- |
| 3 | 1 | 2 | 1 | 26750.960(3) | 26641.579(10) | -- |
| 3 | 0 | 2 | 0 | 27440.992(3) | 27338.687(10) | 26710.491(10) |
| 1 | 0 | 2 | 1 | 34436.849(3) | -- | -- |
| 4 | 1 | 3 | 1 | 35701.086(3) | 35556.827(10) | 34767.037(10) |
| 4 | 0 | 3 | 0 | 36406.672(3) | 36263.248(10) | 35451.242(10) |

**Table 2**. Fit Hamiltonian parameters of *trans*-methyl formate and its singly-substituted $^{13}C$ isotopologues.

| Parameter | Operator | Normal Species | *Ab initio*[a] | $H^{13}COOCH_3$ | $HCOO^{13}CH_3$ |
|---|---|---|---|---|---|
| $V_3$ ($cm^{-1}$) | ½ (1-cos3$\gamma$) | 12.955(17) | 22.5 | 12.955[c] | 12.955[c] |
| $F$ ($cm^{-1}$) | $P_\gamma^2$ | 7.04[b] | 7.04 | 7.04[c] | 7.04[c] |
| $\rho$ (unitless) | $P_\gamma P_a$ | 0.27347(3) | 0.274 | 0.26922(6) | 0.27343(7) |
| $A_{RAM}$ (MHz) | $P_a^2$ | 47390(3) | 47908.11 | 46908(6) | 47306(6) |
| $B_{RAM}$ (MHz) | $P_b^2$ | 4784.25(11) | 4740.59 | 4770.339(3) | 4649.5182(26) |
| $C_{RAM}$ (MHz) | $P_c^2$ | 4398.490(4) | 4368.00 | 4380.8047(14) | 4284.0389(13) |
| $D_{ab}$ (MHz) | $P_aP_b + P_bP_a$ | 1250(4) | 1812.44 | 1252.2(15) | 1203.8(14) |
| $d_{ab}$ (MHz) | (1-cos3$\gamma$) $(P_aP_b + P_bP_a)$ | 661(4) | | 661.9(17) | 647.8(17) |
| $G_v$ (MHz) | $P^2P_\gamma^2$ | 1.27(5) | | 1.27[c] | 1.27[c] |
| $\Delta_{ab}$ (MHz) | $P_\gamma^2(P_aP_b + P_bP_a)$ | 15.2(17) | | 15.2[c] | 15.2[c] |
| $\Delta_J$ (kHz) | $-P^4$ | 1.102(9) | | 1.102[c] | 1.102[c] |
| $\Delta_{JK}$ (kHz) | $-P^2P_a^2$ | -117.6(18) | | -117.6[c] | -117.6[c] |
| $\delta_J$ (kHz) | $-2P^2(P_b^2-P_c^2)$ | 0.0531(11) | | 0.0531[c] | 0.0531[c] |
| $N_{lines}$ | | 33 | | 13 | 13 |
| $\sigma_{fit}$[d] | | 0.97 | | 1.02 | 0.90 |

[a]Calculated at an MP2/6-311++G(d,p) level of theory.
[b]Fixed to its *ab initio* value.
[c]Fixed to normal species value.
[d]Unitless standard deviation between observed and calculated frequencies.

**Table 3**. Derived substitution coordinates[a] of the two carbon atoms of *trans*-methyl formate.

| Carbon atom | Substitution Coordinates | | *Ab initio*[b] | |
|---|---|---|---|---|
| | $\|a\|$ | $\|b\|$ | $a$ | $b$ |
| $C_{carbonyl}$ | 0.598(3) | 0.334(5) | 0.599 | 0.356 |
| $C_{methyl}$ | 1.7578(10) | 0.143(12) | -1.778 | 0.149 |

[a]All coordinates are reported in angstroms.
[b]Calculated at an MP2/6-311++G(d,p) level of theory.

**Table 4**. Transitions of *trans*-methyl formate sought toward Sgr B2(N).

| *A* species | | | | | | | | | |
|---|---|---|---|---|---|---|---|---|---|
| Transition | Frequency (MHz) | $S\mu^2$ ($D^2$) | $E_l$ (K) | $\theta_{HPBW}$ (arcsec) | $T_c$ (K) | $\eta_B$ [a] | $V_{LSR}$ (km $s^{-1}$) | $\Delta T_a$* (mK) | $\Delta V$ (km $s^{-1}$) |
| $1_{01}$-$0_{00}$ | 9124.221 | 17.07 | 0 | 81 | 25.6 | 0.978 | 60.7(6) | -11.8(7) | 12.4(12) |
| $2_{12}$-$1_{11}$ | 17921.508 | 25.61 | 3.23 | 41 | 13.5 | 0.919 | -- | -- [b] | -- |
| $2_{02}$-$1_{01}$ | 18247.038 | 34.15 | 0.44 | 41 | 10.7 | 0.915 | 64.5(4) | -20.9(12) | 13.3(9) |
| $2_{11}$-$1_{10}$ | 18575.916 | 25.61 | 3.25 | 40 | 10.9 | 0.912 | 67.8(6) | -11.7(7) | 19.0(12) |
| $3_{13}$-$2_{12}$ | 26881.342 | 45.53 | 4.09 | 28 | 6.8 | 0.825 | -- | -- [b] | -- |
| $3_{03}$-$2_{02}$ | 27367.048 | 51.22 | 1.32 | 27 | 7.1 | 0.819 | 64.9(3) | -21.2(16) | 9.4(8) |
| $5_{05}$-$4_{04}$ | 45593.037 | 85.35 | 4.38 | 16 | 2.7 | 0.574 | -- | -- [c] | -- |

| *E* species | | | | | | | | | |
|---|---|---|---|---|---|---|---|---|---|
| Transition | Frequency (MHz) | $S\mu^2$ ($D^2$) | $E_l$ (K) [d] | $\theta_{HPBW}$ (arcsec) | $T_c$ (K) | $\eta_B$ [a] | $V_{LSR}$ (km $s^{-1}$) | $\Delta T_a$* (mK) | $\Delta V$ (km $s^{-1}$) |
| $1_0$-$0_0$ | 9207.427 | 18.03 | 2.07 | 81 | 25.6 | 0.978 | 64.0(7) | -9.4(6) | 18.1(14) |
| $2_1$-$1_1$ | 17820.962 | 23.66 | 0.00 | 42 | 13.6 | 0.919 | -- | -- [b] | -- |
| $2_0$-$1_0$ | 18367.848 | 35.65 | 2.51 | 40 | 10.7 | 0.914 | 62.9(12) | -19.2(7) | 7.6(12) |
| $3_1$-$2_1$ | 26750.960 | 42.15 | 0.86 | 28 | 6.8 | 0.826 | -- | -- [b] | -- |
| $3_0$-$2_0$ | 27440.992 | 52.56 | 3.39 | 27 | 7.1 | 0.818 | 70.3(5) | -22.7(10) | 24.3(14) |

[a] Assuming a surface error of 390 μm.
[b] Not conclusively detected (see Figure 7).
[c] Not detected (8 mK rms).
[d] Relative to the lowest-energy *E* species rotational level ($1_1$).

**Figure Captions**

**Figure 1**. Conformational potential energy surface of methyl formate, calculated at an MP2/6-31++G(d,p) level of theory.

**Figure 2**. CP-FTMW spectra of methyl formate. Panel A: Spectrum with no discharge (31,000 signal averages, 90 minutes data collection time) showing the relative abundances of *cis*- and *trans*-methyl formate. Panel B: Comparison of spectra with and without a discharge nozzle.

**Figure 3**. Laboratory spectrum of *trans*-methyl formate from Figure 2 (in red) overlaid with the spectrum observed toward Sgr B2(N) (in black) in the region of the $J$ = 1-0, $K_a$ = 0 transitions. The interstellar spectrum has been Doppler shifted to the +64 km $s^{-1}$ velocity of the Sgr B2(N) core. The intensity scale is for the interstellar spectrum (the intensities of the laboratory transitions are given in Figure 2).

**Figure 4**. Plot of $K_a$ = 0 transitions of *trans*-methyl formate sought in Sgr B2(N). The strong emission features seen near the $2_{01}$-$1_{01}$ ($A$), $3_{03}$-$2_{02}$ ($A$), and $3_0$-$2_0$ ($E$) transitions are due to vibrationally excited states of $HC_3N$. The dashed line indicates the +64 km $s^{-1}$ velocity characteristic of Sgr B2(N).

**Figure 5**. Excitation temperature determination of *trans*-methyl formate in Sgr B2(N). Panel A: Observations of the $2_{02}$-$1_{01}$ and $2_{11}$-$1_{10}$ $A$ species transitions. Panel B: Relative intensity of these two transitions as a function of excitation temperature. The red horizontal line indicates the value observed in Sgr B2(N), while the blue lines indicates the error bars.

**Figure 6**. Plot of six Sgr B2(N) lines ($1_{01}$-$0_{00}$ (*A*), $2_{02}$-$1_{01}$ (*A*), $2_{11}$-$1_{10}$ (*A*), $1_0$-$0_0$ (*E*), $2_0$-$1_0$ (*E*), and $3_0$-$2_0$ (*E*)) averaged together. The three dashed lines indicate +64, +73, and +82 km $s^{-1}$ LSR velocities.

**Figure 7.** Observations of four tentatively detected $K = 1$ transitions of *trans*-methyl formate.

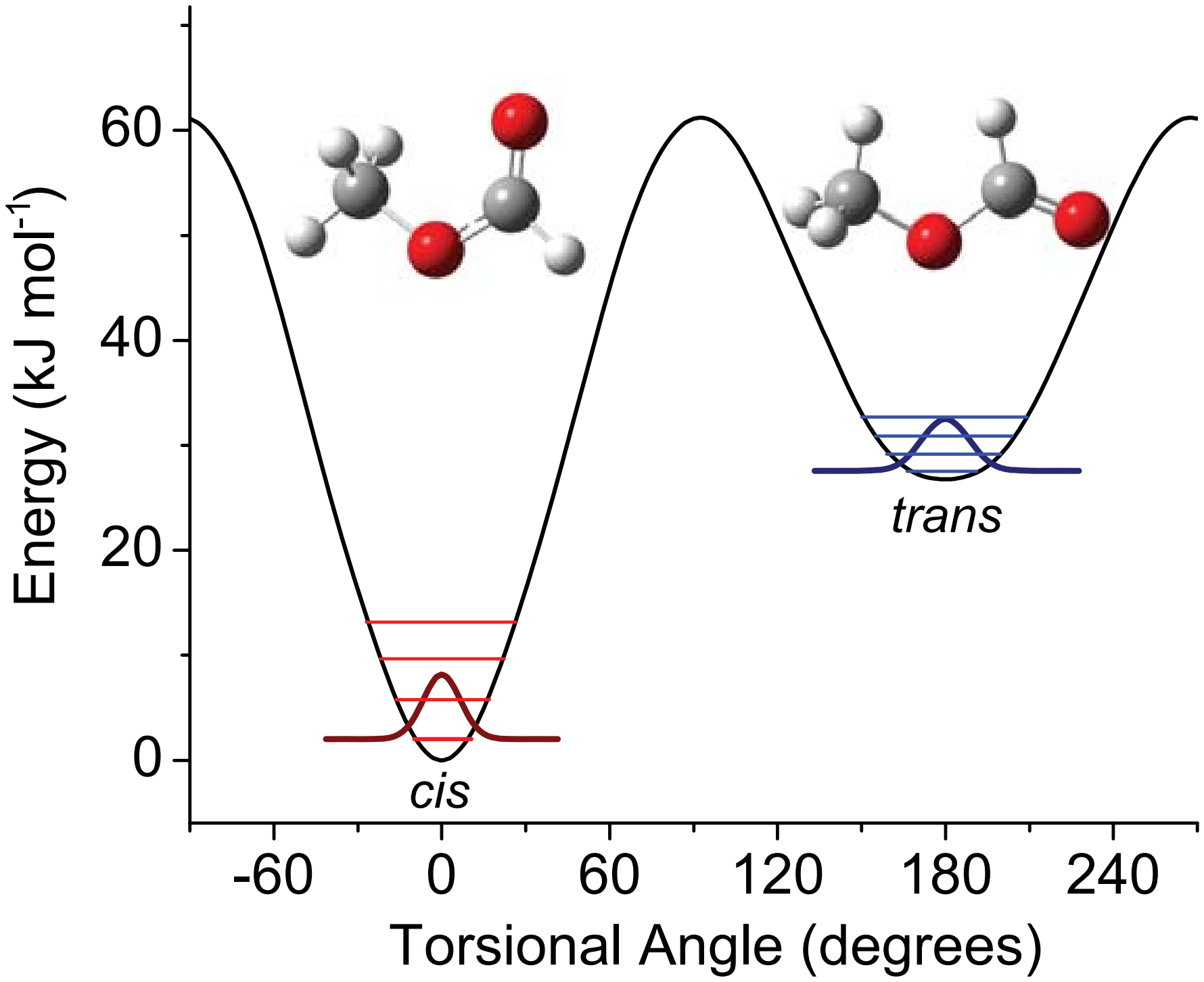

Figure 1

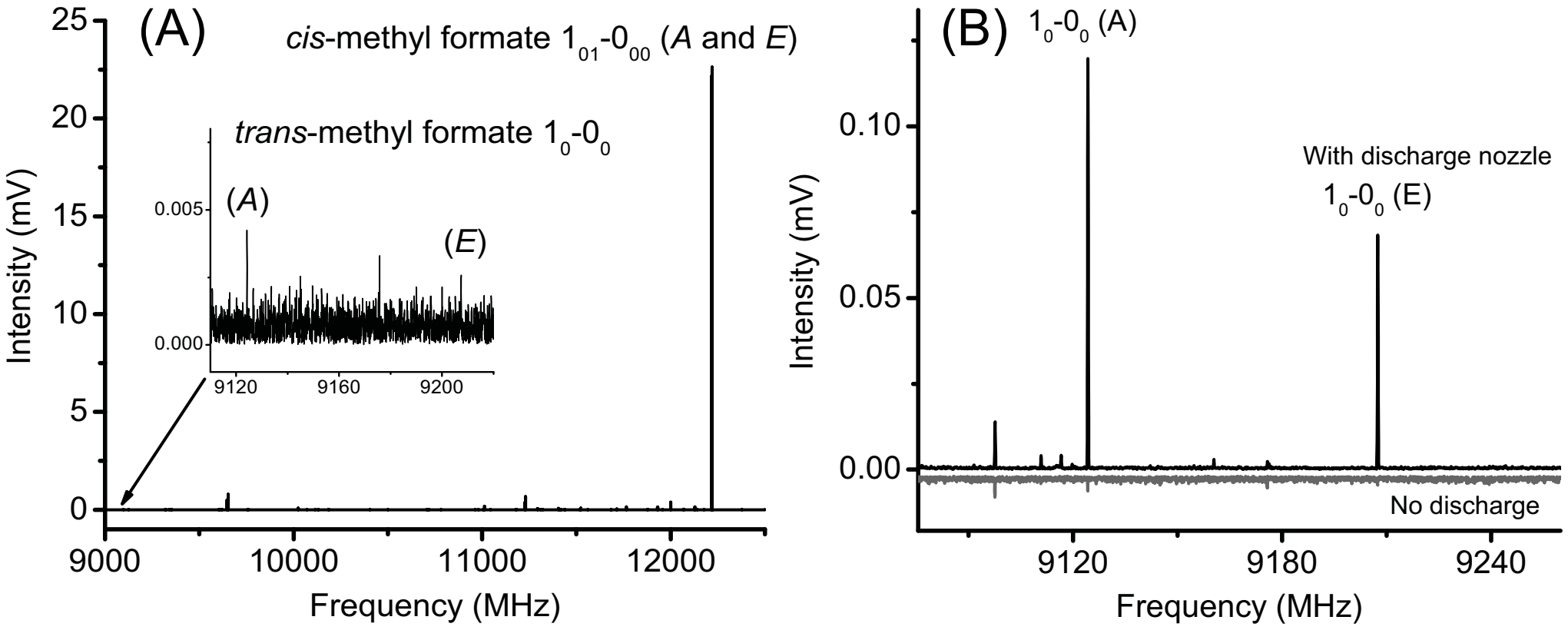

Figure 2

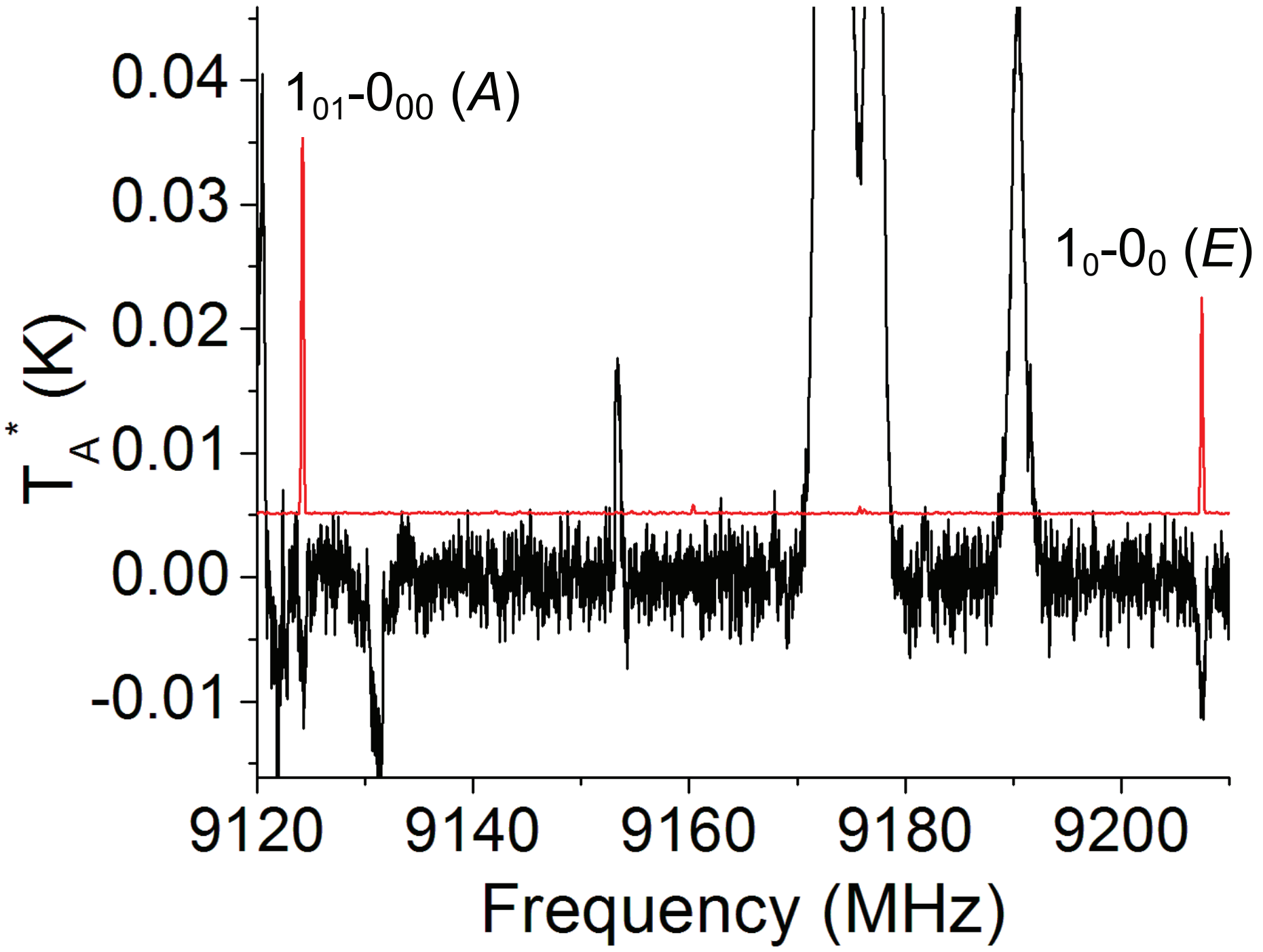

Figure 3

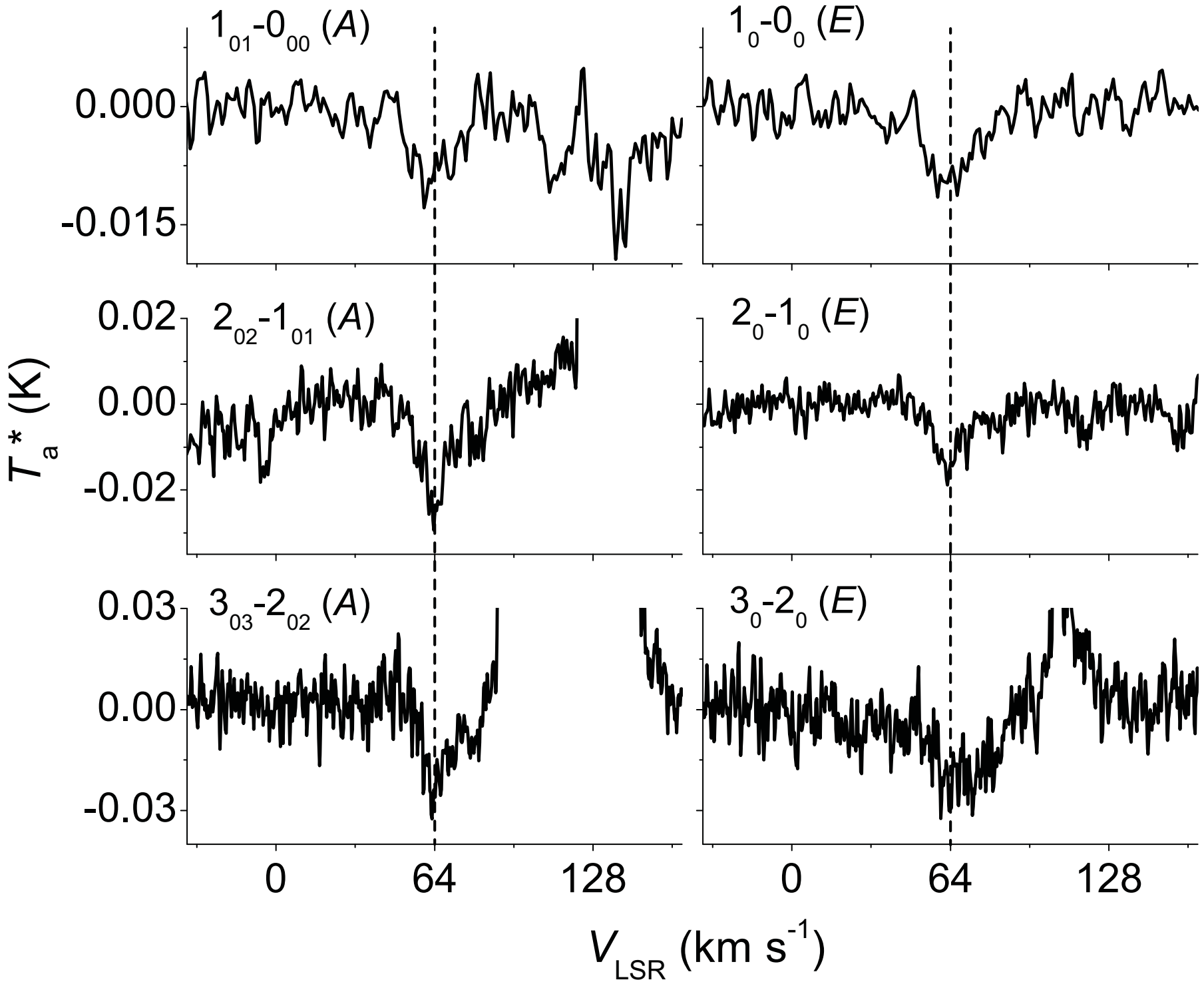

Figure 4

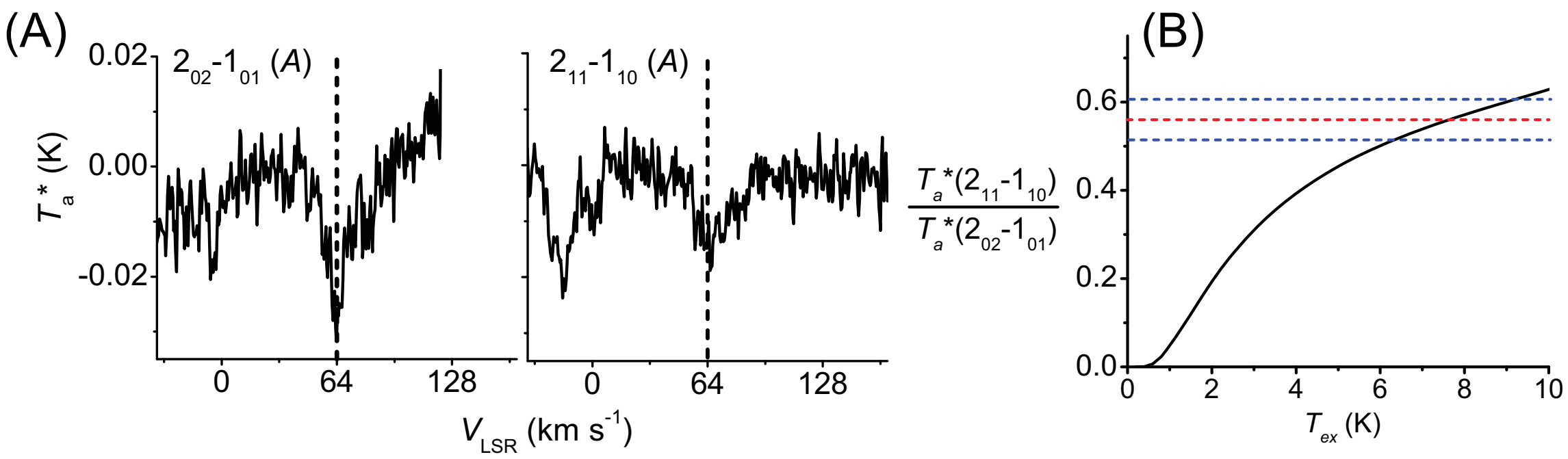

(A)
0.02
0.00
-0.02
$T_a^*$ (K)
$2_{02}$-$1_{01}$ (*A*)
$2_{11}$-$1_{10}$ (*A*)
0
64
128
$V_{LSR}$ (km s$^{-1}$)
$\frac{T_a^*(2_{11}\text{-}1_{10})}{T_a^*(2_{02}\text{-}1_{01})}$
(B)
0.0
0.2
0.4
0.6
0
2
4
6
8
10
$T_{ex}$ (K)

Figure 5

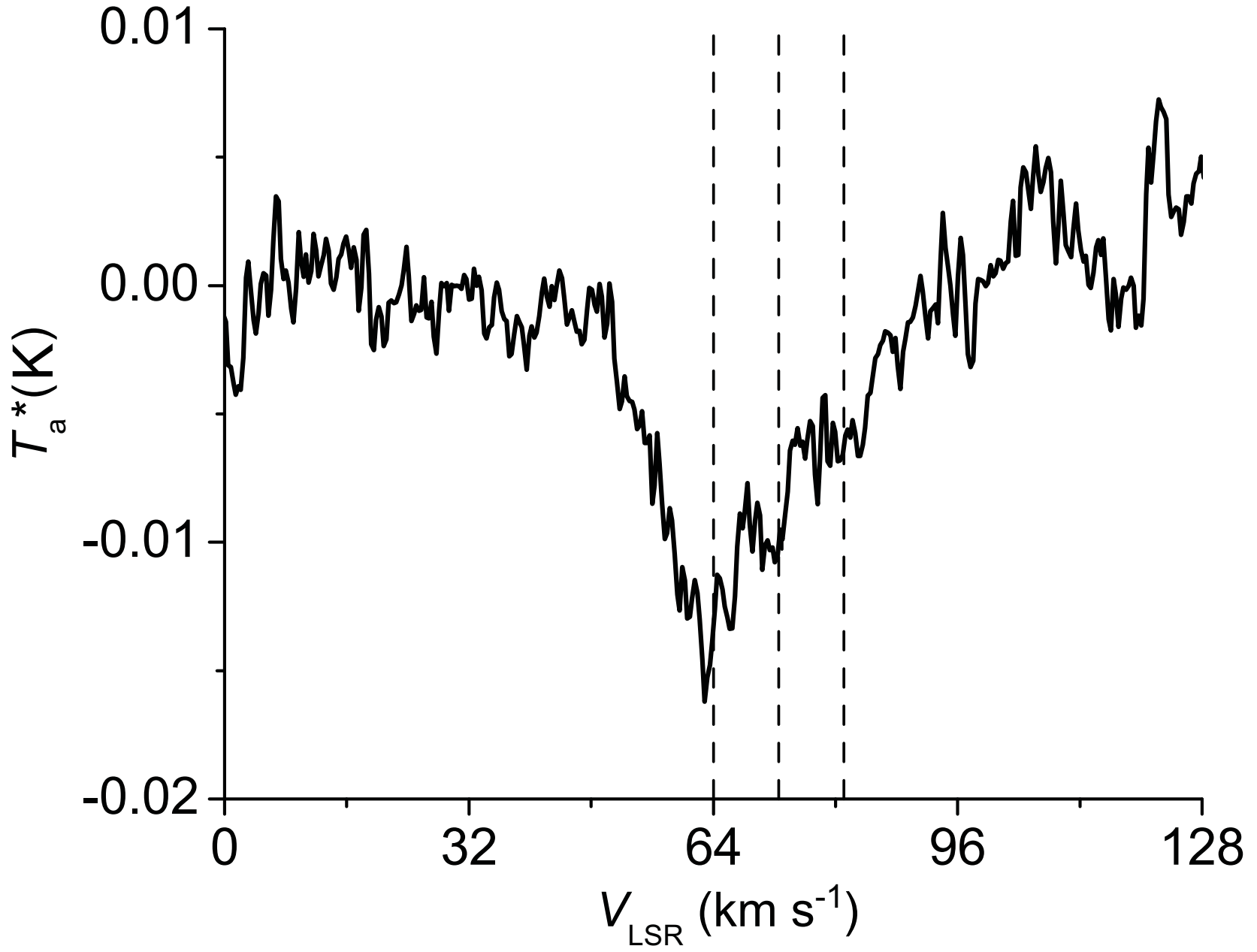

Figure 6

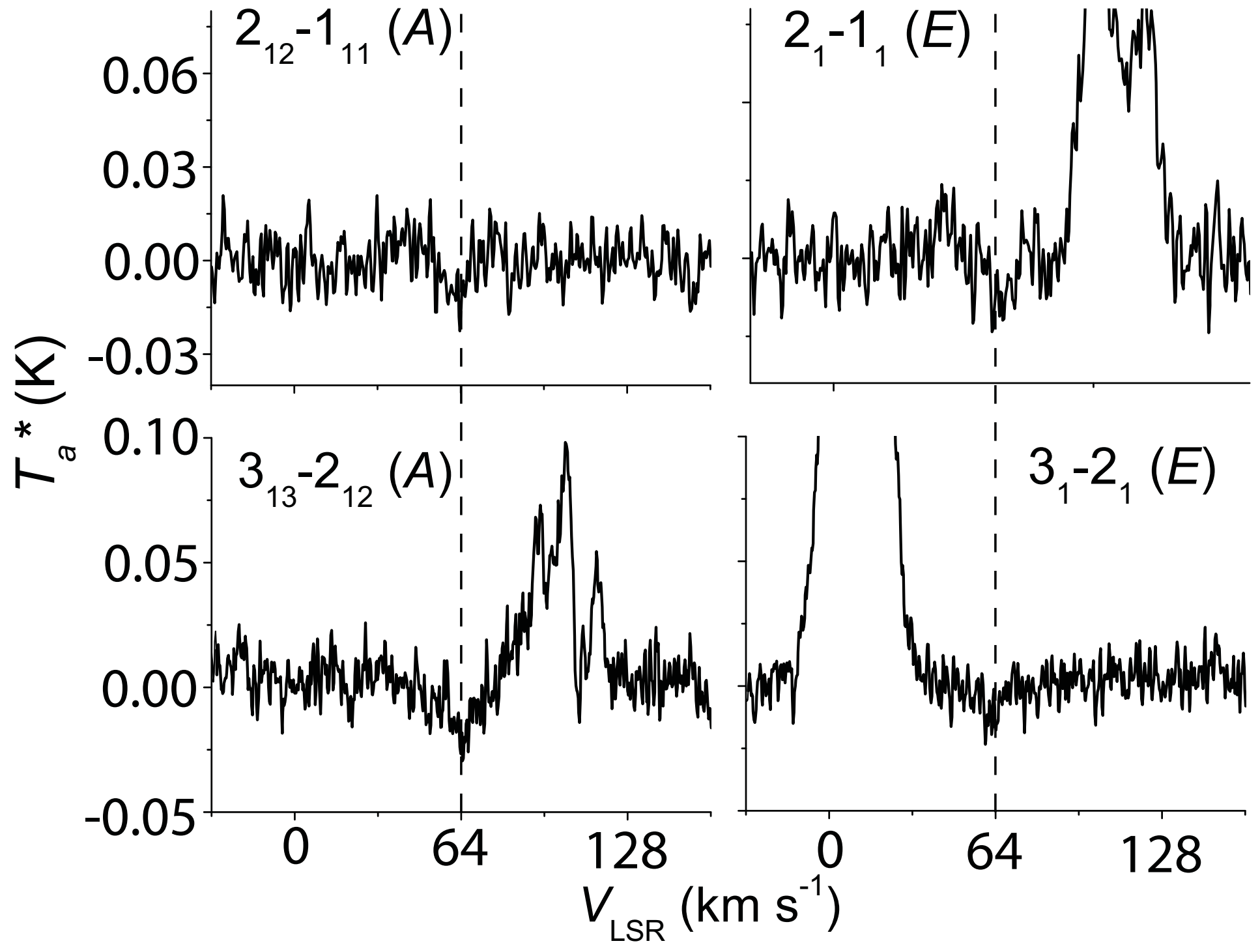

Figure 7